# A Predictor-Corrector Method for Power System Variable Step Numerical Simulation

Yiming Cai, *Student Member, IEEE*, Junbo Zhang, *Senior Member, IEEE*, and Weizhou Yu, *Student Member, IEEE*

***Abstract*-- This letter proposes a predictor-corrector method to strike a balance between simulation accuracy and efficiency by appropriately tuning the numerical integration step length of a power system time-domain simulation. Numerical tests indicate that, by estimating the truncation error for step length tuning based on the 2-Step Adams-Moulton method and the implicit Trapezoidal method, the proposed method can provide much more precise results at little cost of efficiency compared to a conventional variable step method based on Newton's method.**

## I. Introduction

THE implicit trapezoidal method (ITM) is widely used in power system time-domain simulations [1]. In most cases, the ITM is implemented with a fixed step, but for long-term simulations, to enhance the computation efficiency, an ITM with variable steps is preferable [2].

In general, a variable step ITM (VITM) can be achieved by tuning the step length according to the indices referring to the consistency and convergence of the numerical calculation. However, in practice, very few indices meet these requirements. For instance, the number of iterations in Newton's method that the ITM takes to solve the nonlinear algebraic equations of the power system model is widely employed to tune the step length of the VITM [3]. The ITM takes care of the numerical calculation convergence but fails to consider the consistency and results in a loss of simulation accuracy. To consider both simulation accuracy and efficiency, this letter proposes a new implicit variable-step method using the predictor-corrector method (PCM), which takes advantage of the A-stable ITM and other implicit methods' high accuracy. Consequently, fast calculations without loss of result accuracy is obtained.

This work was supported in part by the Natural Science Foundation of Guangdong Province, China (2018B030306041), the National Natural Science Foundation of China (51607071), and the Fundamental Research Funds for the Central Universities (2017JQ011).

Yiming Cai, Junbo Zhang (Corresponding author) and Weizhou Yu are with School of Electrical Power, South China University of Technology, Guangzhou, Guangdong, P. R. China (e-mail: yimcai@student.ethz.ch, epjbzhang@scut.edu.cn, yu.weizhou@mail.scut.edu.cn). Yiming Cai is also with Department of Information Technology and Electrical Engineering, ETH Zürich.

## II. A Brief Review to the Numerical Algorithms

### A. *Power System Time-domain Simulation Model*

In power system time-domain simulations, the system is represented with differential-algebraic equations (DAEs) [2]:

$$x' = f(x, y, t), x(t_0) = x_0 \tag{1}$$

$$0 = g(x, y, t), y(t_0) = y_0 \tag{2}$$

where $x$ denotes the state variables; $y$ denotes the algebraic variables; $t$ denotes the time; $f$ and $g$ are the differential and algebraic functions, respectively; and the subscript 0 denotes the start time.

### B. *Numerical Algorithms*

To solve the DAEs, (1) is transferred to a difference form, and with (2), a set of nonlinear algebraic equations is obtained and can be solved using Newton's method [3]. Generally, different numerical methods can be applied for the transformation, but the consistency and convergence are still the key factors that determine the employment of a numerical method. In this letter, the ITM and the $k$-step Adams-Moulton method (AM-$k$, normally with $k$ = 2) are used for power system simulations.

In ITM, (1) is reformed using the Trapezoidal equation:

$$x_{n+1} = x_n + h[f(x_{n+1}, t_{n+1}) + f(x_n, t_n)]/2 \tag{3}$$

where $n$ is the discrete time and $h$ denotes the step length of the calculation. In addition, in the AM-2, (1) is reformed using an order-3 difference equation:

$$x_{n+1} = x_n + h[5f_{n+1} + 8f_n - f_{n-1}]/12 \tag{4}$$

where $f(x_n, t_n)$ is marked $f_n$ for simplicity. Basically, ITM and AM-2 have local truncation errors of $O(h^2)$ and $O(h^3)$ respectively, which means AM-2 is one order more accurate than ITM. On the other hand, ITM is A-stable, while AM-2 is stable with a constraint of

$$-6 < \alpha h < 0 \tag{5}$$

$$\alpha = \mathrm{Re}(\lambda) \tag{6}$$

with $\mathrm{Re}(\lambda)$ the real part of $\lambda$, and $\lambda$ the system's eigenvalue. When applying AM-2 in power system long term simulations, accumulative error is a major concern. Fortunately, we will show in the later that AM-2 can work stably.

### C. Conventional VITM

The conventional VITM is an ITM with a variable step length determined by an index according to the number of iterations that Newton's method takes to solve the nonlinear algebraic equations [3]. A simple instruction during each step of the ITM could be to multiply the step length $h$ with different coefficients according to the number of iterations in Newton's method. If the number of iterations is smaller than a minimum threshold, e.g., 10, the result converges well, and the coefficient could be 1.3. If the number of iterations is larger than a maximum threshold, e.g., 15, the results do not converge well, and the coefficient could be 0.9. For other situations, the step length remains unchanged. The final step size is limited to an interval, e.g., [0.01, 0.16].

The above measure takes care of the numerical calculation convergence but fails to consider the consistency. In many cases, Newton's method converges rapidly; therefore, the step length of the VITM will increase rapidly, but the simulation accuracy cannot be guaranteed.

## III. Proposed Variable Step Method

To consider consistency and convergence of the numerical calculation simultaneously, the step length of the VITM can be tuned by estimating the truncation error $G_{n+1}$ at each step using a PCM [4], so the variation of the step length can be appropriately controlled, and the loss of simulation accuracy can be consequently avoided.

In the PCM, we assume two numerical methods, P and Q, are of accuracy of order $p$ and $q$ ($p<q$), respectively, so the local truncation errors of P and Q are $O(h^{p+1})$ and $O(h^{q+1})$, respectively. When P is applied at the step $n$+1, the truncation error $T_{n+1(p)}$ can be presented as [1]:

$$T_{n+1(p)} = x_{n+1} - x_{n+1(p)} = K_1 h^{p+1} x_t^{p+1} + O(h^{p+2}) \quad (7)$$

where $x_{n+1}$ is the precise value at the step $n$+1, $x_{n+1(p)}$ is the predicted value with P, $K_1$ is a constant, and $x_t^{p+1}$ is the ($p$+1) derivative of $x_t$ at moment $n$. Then, Q is applied to obtain a corrector of $x_{n+1}$. The correcting phase can be repeated to obtain new correctors with old ones to improve precision of the calculation. If the correcting phase is repeated for $m$ times, the truncation error $T_{n+1(q)}$ would be:

$$T_{n+1(q)} = x_{n+1} - x_{n+1(q)} = K_2 h^{p+m+1} x_t^{p+m+1} + O(h^{p+m+2}) \quad (8)$$

Since $p+m \leq q$ and $m \geqslant 1$, subtracting (7) from (8), we have

$$x_{n+1(q)} - x_{n+1(p)} = K_1 h^{p+1} x_t^{p+1} + O(h^{p+2}) \quad (9)$$

Comparing (9) with (7), we have:

$$T_{n+1(p)} = x_{n+1(q)} - x_{n+1(p)} + O(h^{p+2}) \quad (10)$$

Clearly, according to (10), if the corrector has higher order than the predictor, the truncation error could be estimated quite accurately with the predictor and the corrector.

Let ITM be P and let AM-2 be Q. According to (10), the truncation error $G_{n+1}$ can be obtained as follows:

$$x_{n+1(AM-2)} = x_{n+1(ITM)} + h[5f_{n+1} + 8f_n - f_{n-1}]/12 \quad (11)$$

$$G_{n+1} = x_{n+1(AM-2)} - x_{n+1(ITM)} \quad (12)$$

It should be noted that AM-2 requires the data at the time stamps $n$ and $n$-1 to calculate the data at the time stamp $n$+1; therefore, there would not be a corrector at the initial step.

The implementation procedure of the proposed method is summarized as follows:

1) For each step, use ITM for the predictor $x_{n+1(ITM)}$ and AM-2 for the corrector $x_{n+1(AM-2)}$, and then estimate the truncation error $G_{n+1}$.
2) If the maximum $G_{n+1}$ is smaller than a minimum threshold, e.g., 5e-5, the predictor is precise enough, and its step length could be doubled.
3) If the maximum $G_{n+1}$ is larger than a maximum threshold, e.g., 5e-4, the predictor is not precise, and the step length should be halved.
4) For other situations, the step length remains unchanged.
5) The final step size is limited in interval, e.g., [0.01, 0.16].

Note that, thresholds are tuned according to the balance of accuracy and efficiency. If accuracy overweighs efficiency in some cases, thresholds should be reduced.

It should be also noticed that only the ITM predictors are recorded as simulation results, while AM-2 corrector is only for step tuning. Hence, simulation error of each step can be approximately evaluated to control variable step with algorithm consistency, and there will be no accumulative error from AM-2. By applying PCM with both implicit methods, we take full advantages of AM-2's high accuracy and ITM's grand numerical stability. Consequently, fast calculations without loss of result accuracy is obtained.

## IV. Simulation Tests

In this section, we test the proposed method in two cases. In the first case, reliability and accuracy of AM-2 is tested with an analytic function in order to illustrate its applicability to be the corrector. In the second case, the proposed method is compared with the conventional VITM in an 8-generator 36-node system to illustrate its advantages.

### A. Test with Analytic Function

An analytic function in (13) with four different time constants is used. ITM and AM-2 with fixed-step 0.01 and 0.001 are conducted to obtain the numerical solutions of $x$ when $t$ varies from 0 to 10, respectively. The errors between the numerical results and the analytic solutions are shown in Fig. 1, and the statistical results are shown in Table I.

$$x(t) = e^{-t/10} + e^{-t/1} + e^{-t/0.1} + e^{-t/0.01} \quad (13)$$

These results illustrate both methods are stable and accurate enough to the systems with a time constant from 10 to 0.01 at a step length smaller than 0.01. The results from AM-2 are closer to the analytical solutions compared to those from ITM, which indicate that AM-2 is more precise than ITM and thus is capable of being the corrector in the proposed method. In addition, it should be pointed out that ITM and AM-2 have the same error at the initial point. This is because AM-2 cannot be initiated at the first step, so ITM is used at the first time stamp.

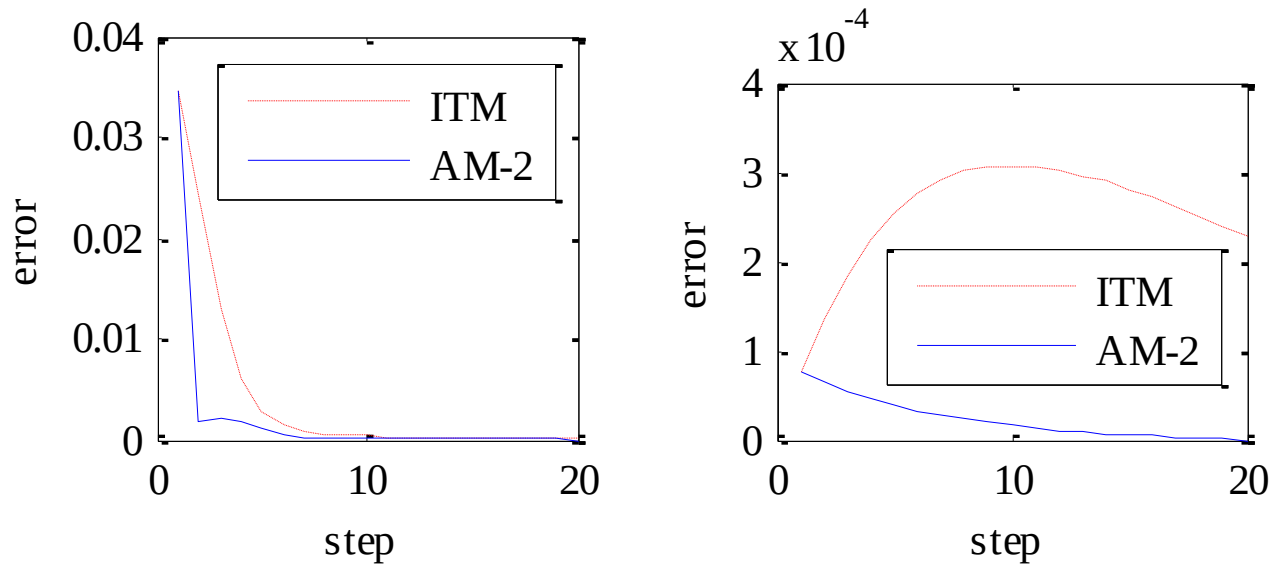

a) Step length 0.01 b) Step length 0.001

Fig. 1. Errors between the numerical results and the analytic solutions.

TABLE I
STATISTICS OF ERRORS OF THE NUMERICAL SOLUTIONS

| Step Length | Method | Max error | Average error |
|---|---|---|---|
| 0.01 | ITM | 0.0346 | 9.1144e-05 |
| | AM-2 | 0.0346 | 4.2491e-05 |
| 0.001 | ITM | 3.0700e-04 | 9.1720e-06 |
| | AM-2 | 7.6000e-05 | 5.1100e-07 |

### B. *Test in the 36-node System*

The 8-generator 36-node system is widely used in power system stability analysis and control. The system diagram and detailed settings are in the Appendix. To enhance complexity of the system, we applied Synchronous Machine VI model, Single Cage Rotor model, three types of Automatic Voltage Regulator, two types of Turbine Governor, three types of Power System Stabilizer, and a Doubly Fed Induction Generator in the original system, which covers most parameter settings in real power systems and shall suffice for the tests.

In this case, statistical tests were carried out to compare performance of the following five different methods in both computation accuracy and efficiency. For the variable step methods, the step length was limited between 0.01 s and 0.16 s.

1) Fixed-step ITM (FITM) with step length 0.01 s;
2) Fixed-step AM-2 (FAM-2) with step length 0.01 s;
3) VITM;
4) Variable-step AM-2 (VAM-2) which uses the number of iterations that AM-2 takes to solve the nonlinear algebraic equations as the index to tune the step length;
5) Proposed predictor-corrector variable-step method.

Various three-phase ground faults were set at 25 of the 36 nodes. The simulation lasted 10 s, and the fault was set at time 1.0 s and was cleared 0.05 s, 0.1 s, 0.15 s and 0.2 s later. In total there were 100 trials.

The bus voltage (including amplitude V and phase angle θ), generator rotor angle δ and speed ω, and the total simulation time were recorded. The test was carried out on a laptop with Intel Core i5-4210 M, CPU 2.6-3.2 GHz and RAM 8 GB. The simulation software was developed with Python 3.6 and was introduced in [5].

The time consumptions for FAM-2 and the proposed method were compared and are shown in Table II. The results illustrate the proposed method is more effective.

The result accuracy for FAM-2 and the proposed method were compared and are shown in Table III, which shows the proposed method can perform similarly to FAM-2, indicating its capability to maintain simulation accuracy.

By setting the FITM as a reference, performances of other methods were compared to those of FITM, and the results are summarized in Table IV.

It is illustrated in column 3 that conventional VITM can effectively improve the computation efficiency more than FITM, but it also introduces large errors. The maximum average differences of voltage phase of the 100 fault simulation cases can even reach 0.8426 rad. In column 4, the situation is even worse, as VAM-2 may lead to a voltage phase difference of 0.9173 rad against FITM, which is unacceptable in practice. In column 5, the results are satisfactory. Compared to FITM, the proposed method introduces only slight differences. Though the proposed method is slightly slower than the other variable step methods, it still improves the efficiency of FITM by 36.91% on average.

TABLE II
EFFICIENCY COMPARISON OF FAM-2 AND THE PROPOSED METHOD

| Method | Shortest time/s | Longest time/s | Average time/s |
|---|---|---|---|
| Proposed | 0.3453 | 16.8772 | 2.7026 |
| FAM-2 | 0.9208 | 18.0723 | 4.2838 |

TABLE III
ACCURACY COMPARISON OF FAM-2 AND THE PROPOSED METHOD

| | Maximum Differences | Average Differences | Variance of Differences |
|---|---|---|---|
| δ (rad) | 0.0025 | 6.52e-04 | 7.0038e-08 |
| ω (p.u.) | 0.0489 | 0.0116 | 2.5661e-05 |
| V (p.u.) | 1.4357 | 0.1145 | 0.0051 |
| θ (rad) | 1.1627 | 0.2300 | 0.0413 |

TABLE IV
COMPARISON OF COMPUTATION ACCURACY AND EFFICIENCY

| Methods | | 1 vs. 2 | 1 vs. 3 | 1 vs. 4 | 1 vs. 5 |
|---|---|---|---|---|---|
| Average Efficiency Improvement | | -0.1% | 62.19% | 54.62% | 36.91% |
| Maximum Average Differences | δ (rad) | 8.31e-04 | 0.0019 | 0.0024 | 4.73e-04 |
| | ω (p.u.) | 0.0127 | 0.0628 | 0.0775 | 0.0105 |
| | V (p.u.) | 0.0995 | 0.5420 | 0.4510 | 0.1295 |
| | θ (rad) | 0.2037 | 0.8426 | 0.9173 | 0.2563 |

Considering both consistency and convergence of the numerical calculation, a compromise of calculation efficiency and accuracy must be made and the proposed method is better than other methods in power system long-term simulations.

## V. CONCLUSION

This letter proposes a predictor-corrector variable step method for power system long-term simulation. Compared to the conventional VITM, the proposed method can consider both consistency and convergence of the numerical calculation by estimating the truncation error for step length tuning, therefore striking a balance between the simulation accuracy and efficiency. Simulations on an analytic function and in the 8-generator 36-node system illustrate the proposed method can achieve accurate results compared to the FITM with a promotion of efficiency by more than 36%.

# APPENDIX

## A. 36-node System Diagram

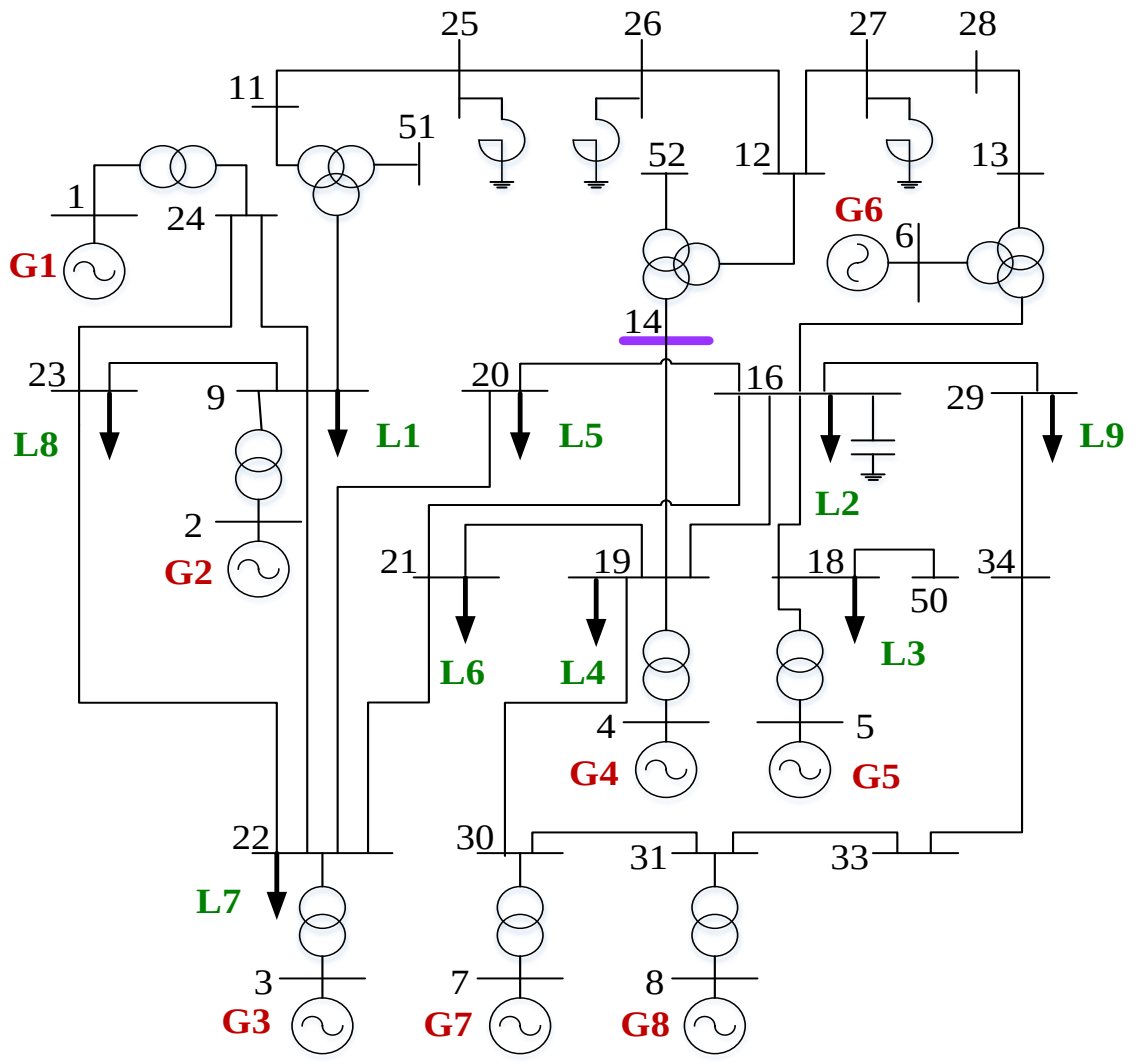

Fig. 2. Diagram of the 36-node system.

## B. Dynamic Models

Dynamic models of the system are listed herein under.

1) Synchronous machine classic VI-order model

TABLE V
PARAMETERS OF SYNCHRONOUS MACHINE

| Variable | Description | Unit |
|---|---|---|
| $S_n$ | Power rating | MVA |
| $V_n$ | Voltage rating | kV |
| $f_n$ | Frequency rating | Hz |
| $x_l$ | Leakage reactance | p.u. |
| $r_a$ | Armature resistance | p.u. |
| $x_d$ | d-axis synchronous reactance | p.u. |
| $x'_d$ | d-axis transient reactance | p.u. |
| $x''_d$ | d-axis sub-transient reactance | p.u. |
| $T'_{d0}$ | d-axis open circuit transient time constant | s |
| $T''_{d0}$ | d-axis open circuit sub-transient time constant | s |
| $X_q$ | q-axis synchronous reactance | p.u. |
| $x'_q$ | q-axis transient reactance | p.u. |
| $x''_q$ | q-axis sub-transient reactance | p.u. |
| $T'_{q0}$ | q-axis open circuit transient time constant | s |
| $T''_{q0}$ | q-axis open circuit sub-transient time constant | s |
| $M = 2H$ | Mechanical starting time (2 × inertia constant) | kWs/kVA |
| D | Damping coefficient | - |
| $K_\omega$ | Speed feedback gain | gain |
| $K_p$ | Active power feedback gain | gain |
| $\gamma_p$ | Active power ratio at node | [0,1] |
| $\gamma_q$ | Reactive power ratio at node | [0,1] |
| $T_{AA}$ | d-axis additional leakage time constant | s |
| $S(1,0)$ | First saturation factor | - |
| $S(2,0)$ | Second saturation factor | - |
| $n_{COI}$ | Center of inertia number | int |
| $u$ | Connection status | {0,1} |

2) Single cage rotor model

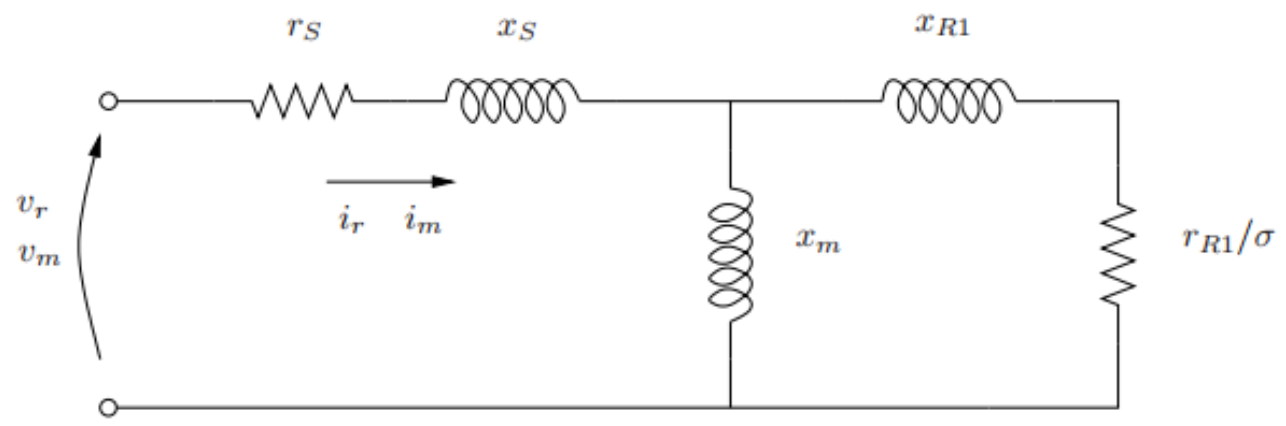

Fig. 3. Electrical circuit of single cage rotor model

TABLE VI
PARAMETERS OF THE SINGLE CAGE ROTOR MODEL

| Variable | Description | Unit |
|---|---|---|
| $H_m$ | Inertia constant | kWs/kVA |
| $r_{R1}$ | 1st cage rotor resistance | p.u. |
| $r_s$ | Stator resistance | p.u. |
| $s_{up}$ | Start-up control | {0,1} |
| $t_{up}$ | Start up time | s |
| $x_{R1}$ | 1st cage rotor reactance | p.u. |
| $x_s$ | Stator reactance | p.u. |
| $x_\mu$ | Magnetization reactance | p.u. |

3) Exciter and Automatic Voltage Regulator

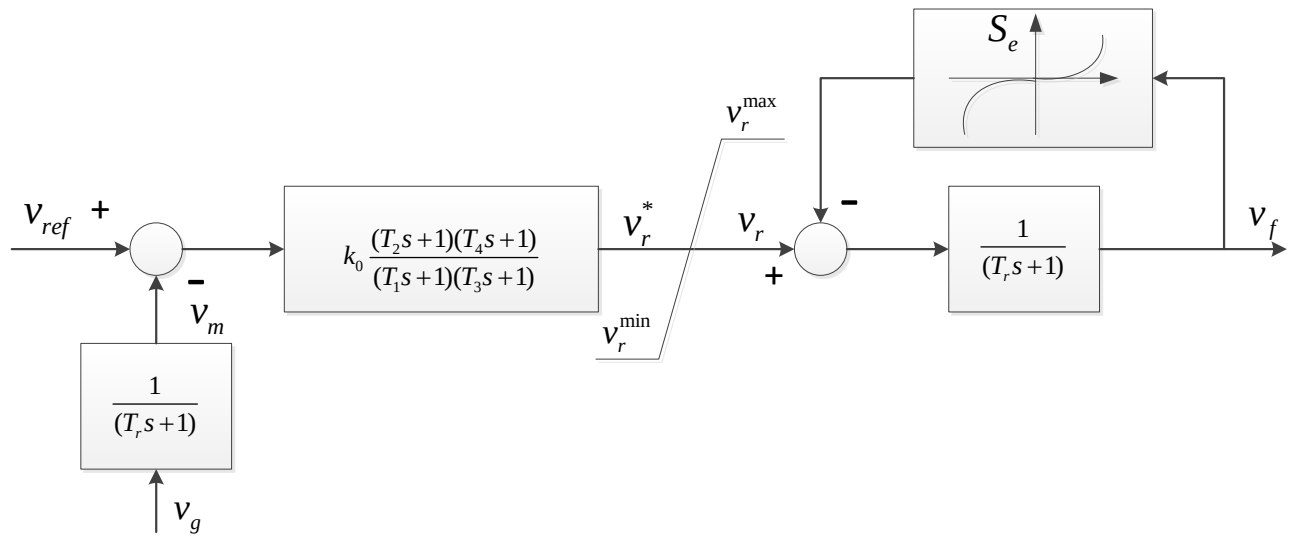

Fig. 4 Exciter Type I

TABLE VII
PARAMETERS OF EXCITER TYPE I

| Variable | Description | Unit |
|---|---|---|
| - | Generator number | int |
| $v_r^{max}$ | Maximum regulator voltage | p.u. |
| $v_r^{min}$ | Minimum regulator voltage | p.u. |
| $K_0$ | Regulator gain | p.u./p.u. |
| $T_1$ | 1st pole | s |
| $T_2$ | 1st zero | s |
| $T_3$ | 2nd pole | s |
| $T_4$ | 2nd zero | s |
| $T_e$ | Field circuit time constant | s |
| $T_r$ | Measurement time constant | s |
| $A_e$ | 1st ceiling coefficient | - |
| $B_e$ | 2nd ceiling coefficient | - |

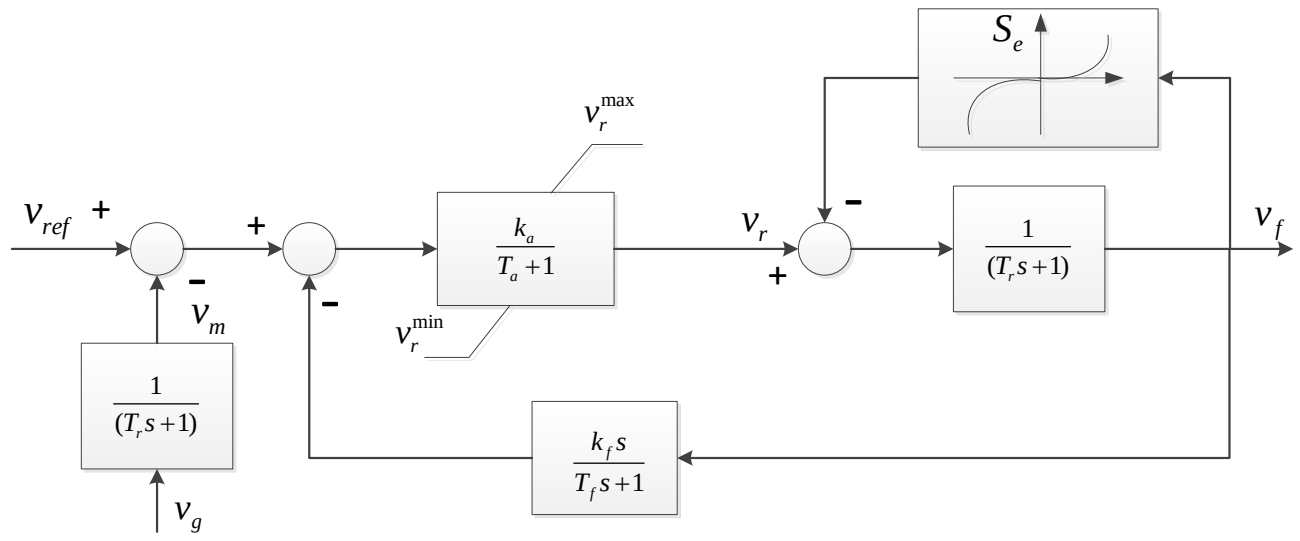

Fig. 5 Exciter Type II

TABLE VIII
PARAMETERS OF EXCITER TYPE II

| Variable | Description | Unit |
|---|---|---|
| - | Generator number | int |
| $v_r^{max}$ | Maximum regulator voltage | p.u. |
| $v_r^{min}$ | Minimum regulator voltage | p.u. |
| $K_a$ | Amplifier gain | p.u./p.u. |
| $T_a$ | Amplifier time constant | s |
| $K_f$ | Stabilizer gain | p.u./p.u. |
| $T_f$ | Stabilizer time constant | s |
| $K_e$ | Field circuit integral deviation | p.u./p.u. |
| $T_e$ | Field circuit time constant | s |
| $T_r$ | Measurement time constant | s |
| $A_e$ | 1st ceiling coefficient | - |
| $B_e$ | 2nd ceiling coefficient | - |

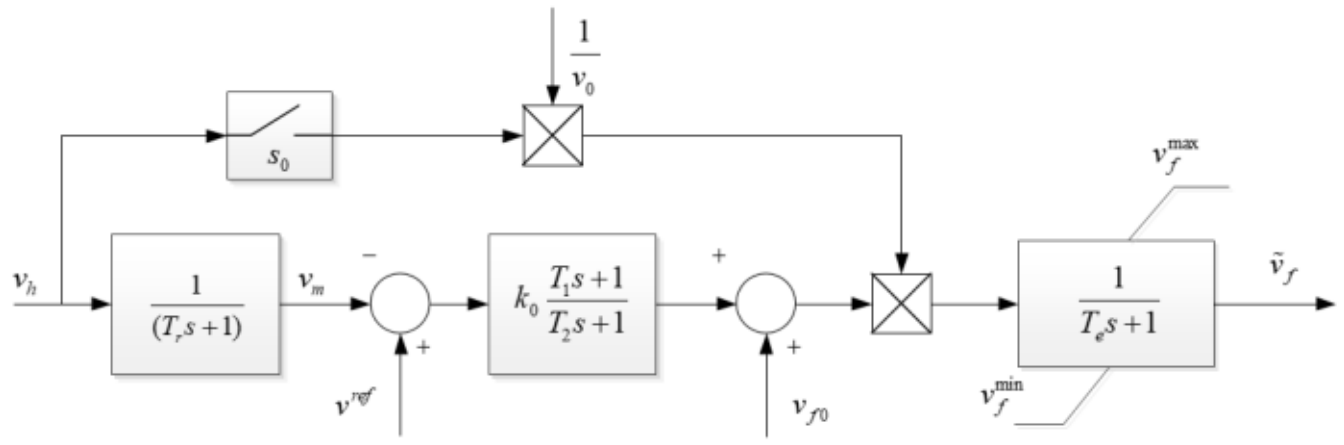

Fig. 6 Exciter Type III

TABLE IX
PARAMETERS OF EXCITER TYPE III

| Variable | Description | Unit |
|---|---|---|
| - | Generator number | int |
| $v_f^{max}$ | Maximum field voltage | p.u. |
| $v_f^{min}$ | Minimum field voltage | p.u. |
| $K_0$ | Regulator gain | p.u./p.u. |
| $T_2$ | Regulator pole | s |
| $T_1$ | Regulator zero | s |
| $v_{f0}$ | Field voltage offset | p.u. |
| $s_0$ | Bus voltage signal | {0,1} |
| $T_e$ | Field circuit time constant | s |
| $T_r$ | Measurement time constant | s |

### 4) Turbine and Governor

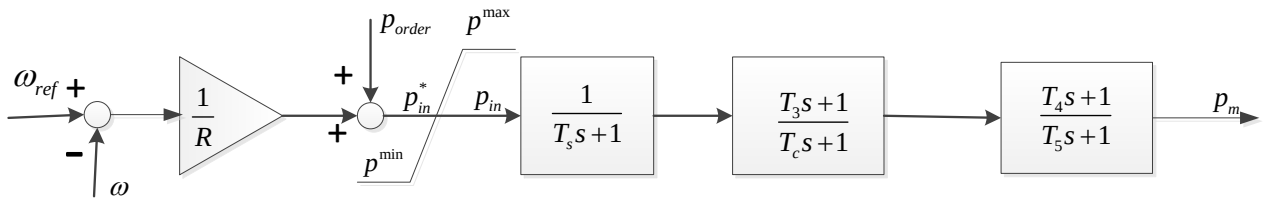

Fig. 7 Turbine governor type I

TABLE X
PARAMETERS OF TURBINE GOVERNOR TYPE I

| Variable | Description | Unit |
|---|---|---|
| $\omega_0^{ref}$ | Reference speed | p.u. |
| $R$ | Droop | p.u. |
| $p^{max}$ | Maximum turbine output | p.u. |
| $p^{min}$ | Minimum turbine output | p.u. |
| $T_S$ | Governor time constant | s |
| $T_C$ | Servo time constant | s |
| $T_3$ | Transient gain time constant | s |
| $T_4$ | Power fraction time constant | s |
| $T_5$ | Reheat time constant | s |

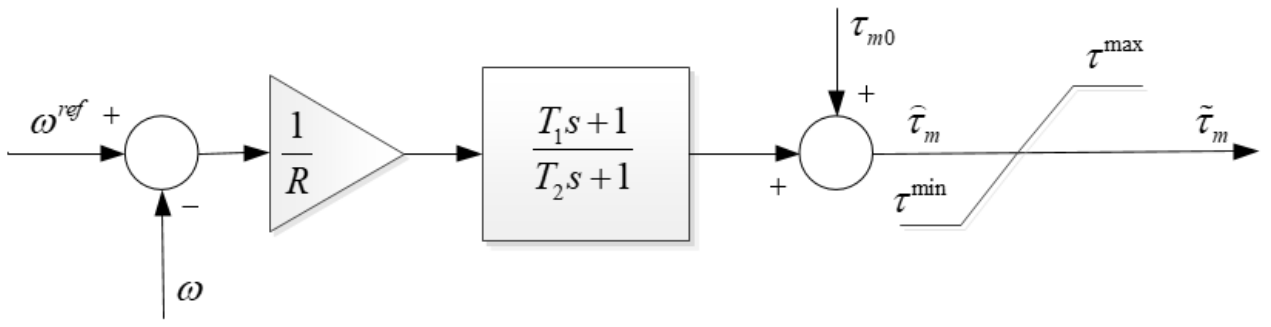

Fig. 8 Turbine governor type II

TABLE XI
PARAMETERS OF TURBINE GOVERNOR TYPE II

| Variable | Description | Unit |
|---|---|---|
| $\omega_0^{ref}$ | Reference speed | p.u. |
| $R$ | Droop | p.u. |
| $p^{max}$ | Maximum turbine output | p.u. |
| $p^{min}$ | Minimum turbine output | p.u. |
| $T_2$ | Governor time constant | s |
| $T_1$ | Servo time constant | s |

### 5) Power System Stabilizer

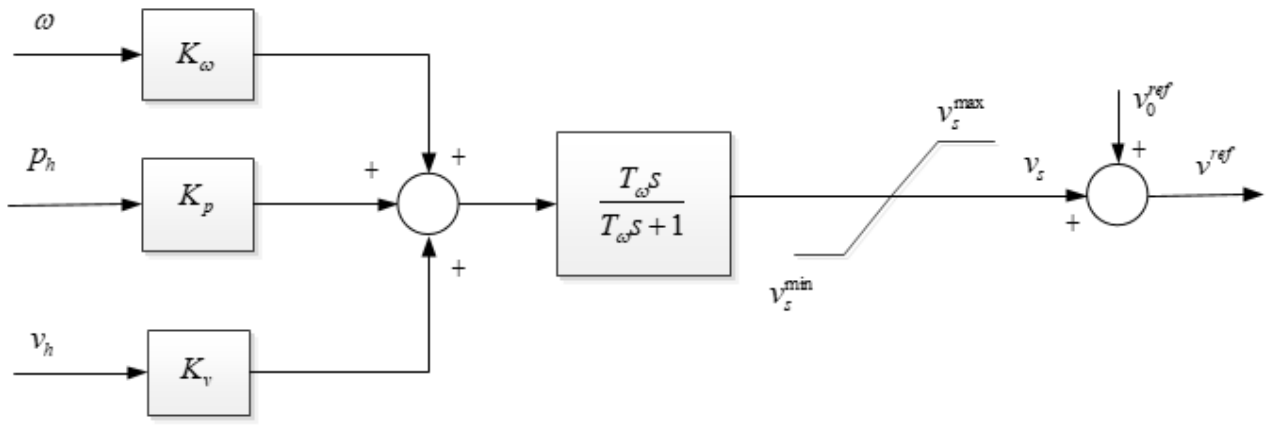

Fig. 9 Power system stabilizer Type I

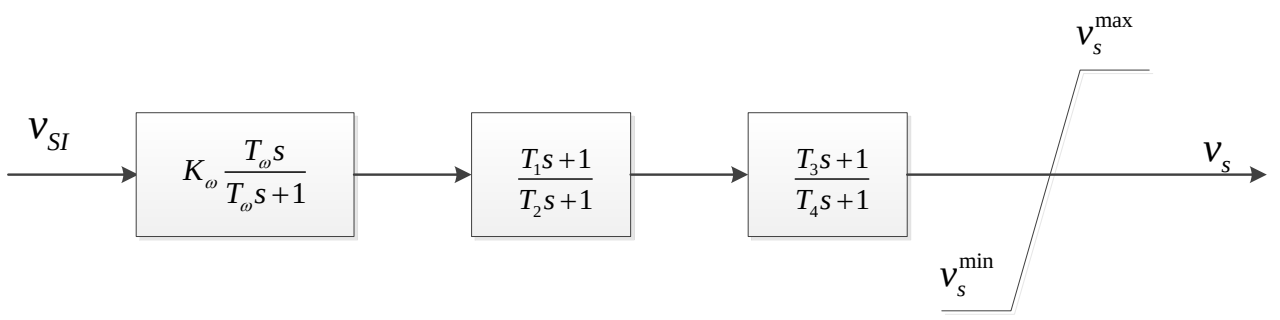

Fig. 10 Power system stabilizer Type II

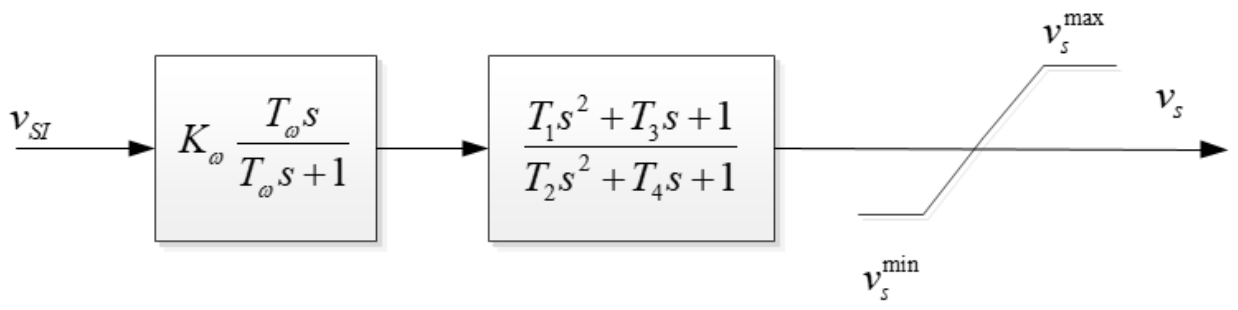

Fig. 11 Power system stabilizer Type III

TABLE XII
PARAMETERS OF POWER SYSTEM STABILIZER

| Variable | Description | Unit |
|---|---|---|
| $v_s^{max}$ | Max stabilizer output signal | p.u. |
| $v_s^{min}$ | Min stabilizer output signal | p.u. |
| $K_\omega$ | Stabilizer gain | p.u./p.u. |
| $T_\omega$ | Wash-out time constant | s |
| $T_1$ | First stabilizer time constant | s |
| $T_2$ | Second stabilizer time constant | s |
| $T_3$ | Third stabilizer time constant | s |
| $T_4$ | Fourth stabilizer time constant | s |
| $K_a$ | Gain for additional signal | p.u./p.u. |
| $T_a$ | Time constant for additional signal | s |
| $K_p$ | Gain for active power | p.u./p.u. |
| $K_v$ | Gain for bus voltage magnitude | p.u./p.u. |

### 6) Doubly Fed Induction Generator

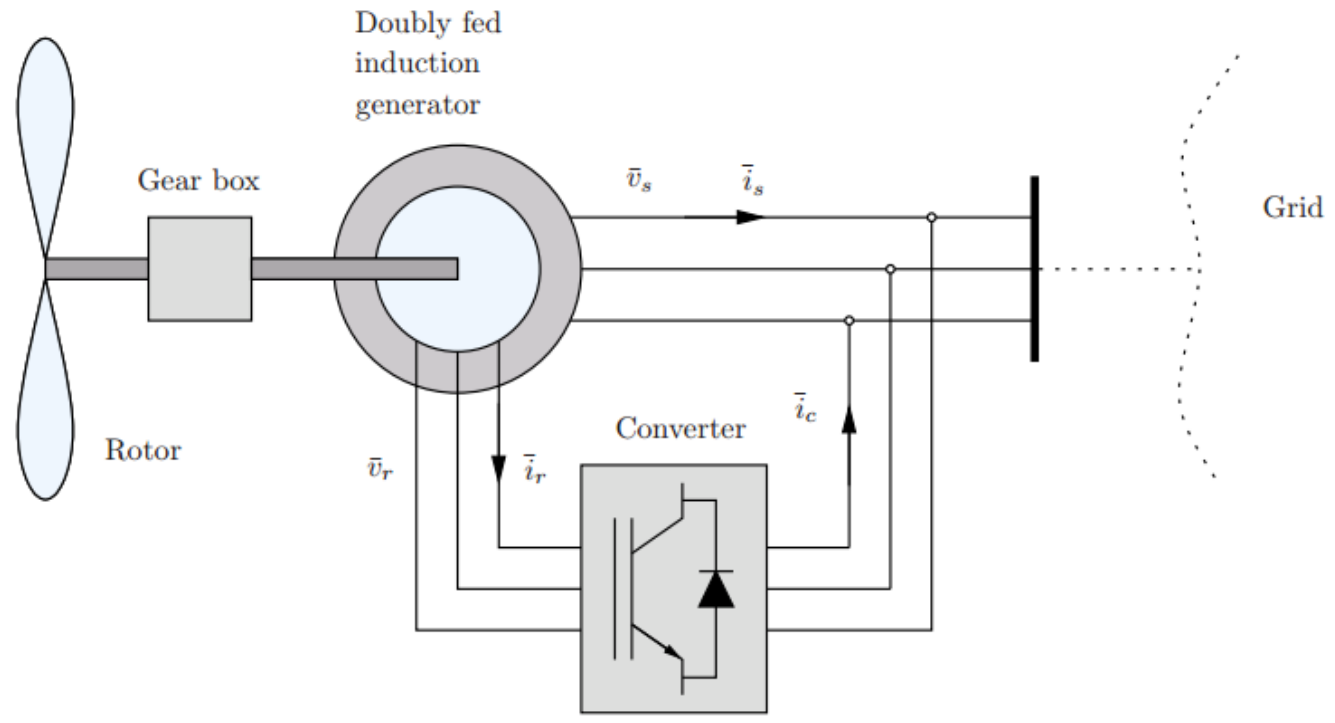

Fig. 12 Doubly Fed Induction Generator

TABLE XIII
PARAMETERS OF WEIBULL'S DISTRIBUTION WIND MODEL

| Variable | Description | Unit |
|---|---|---|
| $v_{\omega n}$ | Nominal wind speed | m/s |
| $\rho$ | Air density | Kg/m$^3$ |
| $\tau$ | Filter time constant | s |
| $\Delta t$ | Sample time for wind measurements | s |
| $c$ | Scale factor for Weibull's distribution | - |
| $k$ | Shape factor for Weibull's distribution | - |

TABLE XIII
PARAMETERS OF DOUBLY FED INDUCTION GENERATOR

| Variable | Description | Unit |
|---|---|---|
| $S_n$ | Power rating | MVA |
| $V_n$ | Voltage rating | kV |
| $f_n$ | Frequency rating | Hz |
| $r_s$ | Stator resistance | p.u. |
| $x_s$ | Stator reactance | p.u. |
| $r_r$ | Rotor resistance | p.u. |
| $x_r$ | Rotor reactance | p.u. |
| $x_\mu$ | Magnetizing reactance | p.u. |
| $H_m$ | Rotor inertia | kWs/kVA |
| $K_p$ | Pitch control gain | - |
| $T_p$ | Pitch control time constant | s |
| $K_v$ | Voltage control gain | - |
| $T_\varepsilon$ | Power control time constant | s |
| $R$ | Rotor radius | m |
| $n_p$ | Number of poles | int |
| $n_b$ | Number of blades | int |
| $\eta_{GB}$ | Gear box ratio | - |
| $p^{max}$ | Maximum active power | p.u. |
| $p^{min}$ | Minimum active power | p.u. |
| $q^{max}$ | Maximum reactive power | p.u. |
| $q^{min}$ | Minimum reactive power | p.u. |

### C. System Data

The system data is listed in the following part, with the same format as in the software PSAT [3].

```
#
Bus 1    10.5
Bus 2    20
Bus 3    10.5
Bus 4    15.7
Bus 5    10.5
Bus 6    10.5
Bus 7    10.5
Bus 8    10.5
Bus 9    220
Bus 10  20
Bus 11  500
Bus 12  500
Bus 13  500
Bus 14  220
Bus 15  20
Bus 16  220
Bus 17  20
Bus 18  220
Bus 19  220
Bus 20  220
Bus 21  220
Bus 22  220
Bus 23  220
Bus 24  220
Bus 25  500
Bus 26  500
Bus 27  500
Bus 28  500
Bus 29  220
Bus 30  220
Bus 31  220
Bus 33  220
Bus 34  220
Bus 50  220
Bus 51  10
Bus 52  10
#
PV 2  100  20    6.000  1.05    6  -6  1.1  0.95
PV 3  100  10.5  3.100  1.05    6  -6  1.1  0.95
PV 4  100  15.7  1.600  1.05    6  -6  1.1  0.95
PV 5  100  10.5  4.300  1.05    6  -6  1.1  0.95
PV 6  100  10.5  -0.01  1.05    6  -6  1.1  0.95
PV 7  100  10.5  2.250  1.05    6  -6  1.1  0.95
PV 8  100  10.5  3.060  1.05    6  -6  1.1  0.95
#
PQ 9   100  220   3.760  2.210  1.1  0.95 1
PQ 16  100  220   5.000  2.300  1.1  0.95 1
PQ 18  100  220   4.300  2.200  1.1  0.95 1
PQ 19  100  220   0.864  0.662  1.1  0.95 1
PQ 20  100  220   0.719  0.474  1.1  0.95 1
PQ 21  100  220   0.700  0.500  1.1  0.95 1
PQ 22  100  220   2.265  1.690  1.1  0.95 1
PQ 23  100  220   2.870  1.440  1.1  0.95 1
PQ 29  100  220   5.200  0.100  1.1  0.95 1
#
SW 1  100  10.5  1.00  0  6  -6  1.1  0.95
#
Shunt 25  100  500  50  0  -1.3665
Shunt 26  100  500  50  0  -1.3665
Shunt 27  100  500  50  0  -1.3665
#
Line 11  25  100  500  50  0  0  0.00  0.0001  0.00  0  0  0  0  0
Line 12  26  100  500  50  0  0  0.00  0.0001  0.00  0  0  0  0  0
Line 12  27  100  500  50  0  0  0.00  0.0001  0.00  0  0  0  0  0
Line 13  28  100  500  50  0  0  0.00  0.0001  0.00  0  0  0  0  0
Line 14  19  100  220  50  0  0  0.0034  0.02  0.00  0  0  0  0  0
Line 16  18  100  220  50  0  0  0.0033  0.0333  0.00  0  0  0  0  0
Line 16  19  100  220  50  0  0  0.0578  0.218  0.3774  0  0  0  0  0
Line 16  20  100  220  50  0  0  0.0165  0.0662  0.4706  0  0  0  0  0
Line 16  21  100  220  50  0  0  0.0374  0.178  0.328  0  0  0  0  0
Line 16  29  100  220  50  0  0  0.00  0.0001  0.00  0  0  0  0  0
Line 18  50  100  220  50  0  0  0.00  0.001  0.00  0  0  0  0  0
Line 19  21  100  220  50  0  0  0.0114  0.037  0.00  0  0  0  0  0
Line 19  30  100  220  50  0  0  0.0196  0.0854  0.162  0  0  0  0  0
Line 20  22  100  220  50  0  0  0.0214  0.0859  0.6016  0  0  0  0  0
Line 21  22  100  220  50  0  0  0.015  0.0607  0.4396  0  0  0  0  0
Line 22  23  100  220  50  0  0  0.0537  0.19  0.3306  0  0  0  0  0
Line 23  24  100  220  50  0  0  0.0106  0.074  0.00  0  0  0  0  0
Line 25  26  100  500  50  0  0  0.0033  0.0343  3.7594  0  0  0  0  0
Line 27  28  100  500  50  0  0  0.00245  0.0255  2.79  0  0  0  0  0
Line 29  34  100  220  50  0  0  0.00  0.0001  0.00  0  0  0  0  0
Line 30  31  100  220  50  0  0  0.00  0.0001  0.00  0  0  0  0  0
Line 31  33  100  220  50  0  0  0.00  0.0001  0.00  0  0  0  0  0
Line 33  34  100  220  50  0  0  0.0154  0.158  0.776  0  0  0  0  0
Line 9    22  100  220  50  0  0  0.0559  0.218  0.3908  0  0  0  0  0
Line 9    23  100  220  50  0  0  0.0034  0.0131  0.00  0  0  0  0  0
Line 9    24  100  220  50  0  0  0.0147  0.104  0.00  0  0  0  0  0
Line 1    24  100  10.5  50  0  0.04773  0.00  0.015  0  0.9302  0  0  0  0
Line 2    9    100  20  50  0  0.09091  0.00  0.0217  0  0.9302  0  0  0  0
Line 3    22  100  10.5  50  0  0.04773  0.00  0.0124  0  0.9302  0  0  0  0
Line 4    19  100  15.7  50  0  0.07136  0.00  0.064  0  0.9756  0  0  0  0
Line 5    18  100  10.5  50  0  0.04773  0.00  0.0375  0  0.9302  0  0  0  0
Line 7    30  100  10.5  50  0  0.04773  0.00  0.0438  0  0.9756  0  0  0  0
Line 8    31  100  10.5  50  0  0.04773  0.00  0.0328  0  0.9756  0  0  0  0
Line 6    17  100  10.5  50  0  0.525  0.00  0.0337  0  1.0000  0  0  0  0
Line 13  17  100  500  50  0  25  0.00  0.01  0  0.9756  0  0  0  0
Line 16  17  100  220  50  0  11  0.00  0.001  0  0.9737  0  0  0  0
Line 9    10  100  220  50  0  11  0.00  -0.002  0  1.0000  0  0  0  0
Line 11  10  100  500  50  0  25  0.00  0.018  0  0.9756  0  0  0  0
Line 51  10  100  10  50  0  0.5  0.00  0.001  0  1.0000  0  0  0  0
Line 12  15  100  500  50  0  25  0.00  0.018  0  0.9756  0  0  0  0
Line 14  15  100  220  50  0  11  0.00  -0.002  0  1.0000  0  0  0  0
Line 52  15  100  10  50  0  0.5  0.00  0.0001  0  1.0000  0  0  0  0
#
Syn6 8    388.4 10.5   50 6   0   0   0.75   0.306 0.196 5.95   0.05   0.611 0.611
    0.196 9999  0.05   8.393 0
Syn6 7    286    10.5   50 6   0   0   0.904 0.358 0.252 5.53   0.05   0.64   0.64
    0.252 9999  0.05   7.692 0
Syn6 6    350    10.5   50 6   0   0   1.633 0.197 0.148 6.92   0.11.633 1.633
    0.148 9999  0.22.62   0
Syn6 5    637.5 10.5   50 6   0   0   1.951 0.306 0.198 6.20.11.951 1.951 0.198
    9999  0.56.149 0
Syn6 4    235    15.7   50 6   0   0   1.81   0.284 0.183 6.20.192 1.81   1.81
    0.183 9999  1.89   6.672 0
Syn6 3    882    10.5   50 6   0   0   1.217 0.349 0.25   7.24   0.10.60.60.25
    9999  0.29.014 0
Syn6 2    780  20  50  6   0   0   2.266 0.27   0.168 8.375 0.224 2.266 2.266
    0.168 9999  1.66   4.249 0
```

```
#
Avr3 2    3     5   0 50  0.1  0.08  0.8  1  1  0.01  0  0
Avr3 3    3     5   0 50  0.1  0.08  0.8  1  1  0.01  0  0
Avr2 4    2     5   0 50  0.03  0.04  0.715  1  0.5  0.03  0  0
Avr2 5    2     5   0 50  0.03  0.04  0.715  1  0.5  0.03  0  0
Avr2 6    2     5   0 50  0.03  0.04  0.715  1  0.5  0.03  0  0
Avr1 7  1  3.3  -2.6 20  2  2  2  2  0.02 0.03  0  0
#
Tg2  7  1  1  0.04  1.1  0  0.03  0.5  0  0.02  1
Tg1  5  1  1  0.05  1.1  0  0.03  0.5  0  0.02  1
Tg1  4  1  1  0.05  1.1  0  0.03  0.5  0  0.02  1
Tg1  2  1  1  0.05  1.1  0  0.03  0.5  0  0.02  1
#
Ind3 19  100  230  60  3  0  0  0.18  0.02  0.12  0.001  0.04  3.5  1  0.51  0  0
0.51  0
#
Pss2  1  2  1  0.1  -0.1  8  10  0.8  0.05  0.8  0.05  0  0  0  0  0  0  0  0  0  0  0  1
Pss1  2  1  1  0.1  -0.1  15  10  0.1631  0.0746  0.2531  0.0246  0  0  0  0  0  0
0  0  0  0  0  1
Pss3  3  1  1  0.1  -0.1  28  10  0.2196  0.2021  0.2196  0.2021  0  0.5  0  0
0.045  0.045  0.045  -0.045  1  0.95  0  1
#
Fault 16  100  220  50  1.0  1.05  0.0000  0.0001
#
Dfig 1  1  1200  10.5  60  0.01  0.1  0.01  0.08  3  3  10  3  10  0.01  75  4  3
0.01123596  1  0  0.7  -0.7  60  1
#
Wind 2  15  1.225  4  0.1  20  2  5  15  1  5  15  0  50  0.01  0.2  50
```